\documentclass[preprint]{aastex}

\usepackage{graphicx,multirow,gensymb}
\usepackage{natbib}
\newcommand{\Msun}{\mbox{M$_{\odot}$}}

\newcommand{\MMJ}{\mbox{M/M$_J$}}
\newcommand{\um}{\mbox{$\mu$m}}
\newcommand{\Jybeam}{\mbox{Jy beam$^{-1}$}}

\newcommand{\cmg}{\mbox{cm$^2$ g$^{-1}$}}
\defcitealias{Sadavoy10}{Paper I} 

\title{``Starless'' Super-Jeans Cores in Four Gould Belt Clouds}
\author{Sarah I. Sadavoy\altaffilmark{1,2}, James Di Francesco\altaffilmark{2,1}, Doug Johnstone\altaffilmark{2,1}}
\altaffiltext{1}{Department of Physics and Astronomy, University of Victoria, PO Box 355, STN CSC, Victoria BC CANADA, V8W 3P6}
\altaffiltext{2}{National Research Council Canada, Herzberg Institute of Astrophysics, 5071 West Saanich Road, Victoria BC CANADA, V9E 2E7}
\email{ssadavoy@uvic.ca}

\begin{document}

\begin{abstract}

From a survey of  729 cores based on JCMT/SCUBA data, we present an analysis of 17 candidate starless cores with masses that exceed their stable Jeans masses. We re-examine the classification of these super-Jeans cores using \emph{Spitzer} maps and find that 3 are re-classified as protostellar, 11 have ambiguous emission near the core positions, and 3 appear to be genuinely starless. We suggest the 3 starless and 11 undetermined super-Jeans cores represent excellent targets for future observational and computational study to understand the evolution of dense cores and the process of star formation.

\end{abstract}


\section{Introduction}\label{Intro}

Stars form in dense cores that are supported internally through various mechanisms, such as thermal pressure (e.g., \citealt{Larson03}), magnetic fields (e.g., \citealt{Shu87}), or turbulent support (e.g., \citealt{Myers99}). While these mechanisms all likely play an important role in opposing core self-gravity, their relative significances remain unknown. Understanding the physical properties of dense molecular cloud cores is paramount to understanding star formation since these regions probe the conditions prior to the onset of forming stars.

One well-studied property of dense cores is their observed masses. In particular, it has been noted that dense core mass functions (CMFs), as derived from (sub)millimeter continuum emission for various clouds, have similar shapes to the stellar initial mass function (IMF; e.g., \citealt{Motte98}, \citealt{Ward-T07}, \citealt{Enoch08}, \citealt{Nutter08}). Such similarities support the notion that CMFs are universal distributions and that dense cores determine the mass of their stellar products. Indeed, the relationship between CMFs and the IMF is a key problem in star formation, and reproducing an IMF-like CMF constrains simulations of cloud collapse to dense cores (e.g., \citealt{Clark08}, \citealt{Bate09}, \citealt{Kunz09}).

Unfortunately, we do not understand the processes that govern CMFs. Dense cores, themselves, depend on the balance between pressures from their external environment, internal supports, and self-gravity.  Taking the simple approximation of a thermally-supported core, we can define its critical mass by the corresponding Jeans mass. Since molecular cloud cores are typically cold (T $\approx 10$ K) and small ($\sim$ 0.07 pc; \citealt{difran07}), their Jeans masses are $M_J \approx 1\  \Msun$. Even with low temperatures, thermal pressure is expected to be a dominant support mechanism (\citealt{Larson03}). For example, cores in the Ophiuchus, Orion, and Pipe clouds have typical internal thermal pressures that match the estimated external pressures from the embedding cloud (\citealt{Johnstone00}, \citealt{Johnstone01}, \citealt{Lada08}). Dense cores that balance self-gravity with internal supports should be ``starless'', whereas cores that cannot balance their self-gravity and external pressure should be forming stars (``protostellar''). Therefore, assuming only thermal supports, starless cores with masses much greater than their Jeans masses (super-Jeans) are unexpected.

Following large-scale surveys in the infrared (e.g., \citealt{Evans09}) and submillimeter (e.g., \citealt{difran08}), several studies have identified populations of cores with embedded young stellar objects (YSOs) and without embedded YSOs (e.g., \citealt{Jorgensen07}, \citealt{Hatchell07},  \citealt{Enoch08},  \citealt{Nutter08}, \citealt{Sadavoy10}). In these surveys, some cores with masses $\sim 5-10$ \Msun\ (e.g., \citealt{Enoch08}, \citealt{Sadavoy10}) have been classified as starless. Since observations suggest these cores are cold and dense, these starless super-Jeans cores cannot be supported by thermal pressure alone and represent an interesting sample of objects. Starless super-Jeans cores require non-thermal supports or must be on the cusp of forming stars.

Here, we examine the 17 most super-Jeans cores in the \citet{Sadavoy10} survey of 555 starless cores and determine if they appear genuinely starless or if there is evidence of ongoing star formation. From this analysis, we make a list of interesting objects for future study.

\section{Data}\label{Data}

We selected our sample of candidate starless cores from the analysis in \citet[][hereafter \citetalias{Sadavoy10}]{Sadavoy10}. \citetalias{Sadavoy10} identified cores using the SCUBA Legacy Catalogue (SLC; see \citealt{difran08}). The cores were originally identified using the algorithm Clumpfind (\citealt{Clumpfind}) with minimum flux thresholds of 90 m\Jybeam\ and minimum sizes of $\sim 9.6\arcsec$. We added additional constraints to the cores, i.e., 850 \um\  flux of $>$ 150 m\Jybeam. In addition, each core was visually inspected to remove diffuse objects or artifacts from flat fielding. The SLC had an effective SCUBA beam of $\sim 23\arcsec$ at 850 \um. 

In \citetalias{Sadavoy10}, Spitzer infrared sources were compared to the submillimeter cores to determine which cores were starless and which were protostellar. First, a series of color criteria was imposed on the Spitzer sources. Second, a flux-limited coincidence condition was imposed to associate color-selected Spitzer sources within a SCUBA core. The infrared data for Ophiuchus and Perseus came from the ``cores to disks'' (c2d) program (\citealt{Evans09}), whereas the data for Taurus and Orion were part of independent surveys (Padgett et al. 2010, in prep, Megeath et al. 2010, in prep). See \citetalias{Sadavoy10} for more details on core selection and classification. Table \ref{coreTable} lists the numbers of starless and protostellar cores in each of the four clouds along with the assumed temperatures and cloud distances. 

The main focus of \citetalias{Sadavoy10} was to develop a robust classification technique and then compare the starless and protostellar core populations in different clouds to look for trends with environment. In these comparisons, it was noted that within a given cloud, the starless cores cover similar mass ranges as the protostellar cores. For Ophiuchus, Taurus, and Perseus, the identified cores had masses of a few \Msun\ at most, whereas for Orion, a high mass star forming region, the identified protostellar and starless cores had masses reaching $\sim 100$ \Msun. Protostellar cores can be massive (i.e., $\sim 10$ \Msun), however there were also many massive cores defined as starless and these ``starless'' cores are super-Jeans. In this paper, we will focus on these unusual starless cores for the four well-sampled clouds in \citetalias{Sadavoy10}: Ophiuchus, Taurus, Perseus, and Orion.

\section{Results} \label{Results}

\subsection{Comparison with the Jeans criterion}\label{CompareJ}

To determine if our candidate starless cores are stable against gravitational collapse, we compared them to Jeans-critical spheres\footnote{Here, we are using the Jeans mass as a scale. The critical Bonnor-Ebert mass (\citealt{Bonnor56}, \citealt{Ebert55}), which assumes hydrostatic equilibrium, is smaller than the Jeans mass by a factor of $\sim 1.8$. Thus a core that is super-Jeans will also be a super-critical Bonnor-Ebert sphere.} (for a review, see \citealt{McKee07}). The Jeans criterion is a simple approximation for exploring the critical conditions of a thermally supported, spherical cloud of gas. The maximum mass for a stable thermally-supported sphere of a given size and temperature is:

\begin{equation}
M_J = 1.9 \left(\frac{T_d}{10 \mbox{ K}}\right) \left(\frac{R_J}{0.07 \mbox{ pc}}\right) M_{\odot}\label{jeansEq}
\end{equation}

\noindent where $T$ is the dust temperature of the core and $R_J$ is the Jeans radius. The Jeans mass is dependent on the temperature and pressure (density), i.e., $M_J \propto T^{3/2}\rho^{-1/2}$, but for simplicity, we express the Jeans mass as a function of the observables, temperature and radius, in Equation \ref{jeansEq}. We used the observed core size to estimate the Jeans radius and we assumed the fiducial temperatures as given in Table \ref{coreTable}. Since the super-Jeans starless cores are generally large (i.e., $\gtrsim 35\arcsec$) with respect to the SCUBA 850 \um\ beam, the observed size is fairly similar to the deconvolved size. (Deconvolving the core size would lower the Jeans mass.) In comparison, the equation for deriving the core mass from dust emission is:

\begin{equation}
M = 0.074 \left(\frac{S_{850}}{\mbox{Jy}}\right) \left(\frac{d}{100 \mbox{ pc}}\right)^2 \left(\frac{\kappa_{850}}{0.01\ \cmg}\right)^{-1} \left[\exp\left(\frac{17 \mbox{ K}}{T}\right) - 1\right] M_{\odot}\label{massEq}
\end{equation}

\noindent where $S_{850}$ is the total 850 \um\ flux, $d$ is the distance, $\kappa_{850}$ is the dust opacity at 850 \um, and $T$ is the dust temperature. For all cores, we assume $\kappa_{850} = 0.01$ \cmg, though $\kappa_{850}$ can vary by a factor of two (\citealt{Henning95}). The adopted distances and temperatures are given in Table \ref{coreTable}.

To first order, we can determine which of our cores are thermally unstable by comparing the observed mass to the expected Jeans mass. Thermally balanced cores should have $\MMJ \lesssim 1$. When a thermally-supported spherical mass exceeds the Jeans condition ($\MMJ > 1$), the sphere should become unstable and should form a protostar within a free-fall time. We adopted super-Jeans limits of  $\MMJ \ge 2$ for Ophiuchus and Taurus and $\MMJ \ge 4.5$ for Perseus and Orion. We chose these limits to identify the starless cores that seemed most unusual relative to the protostellar cores. 
 
Mass estimates from submillimeter dust emission are only accurate within a factor of a few. Nevertheless, we are examining cores that are unusual relative to the other cores in the given cloud and are not comparing individual \MMJ\ ratios (hence the different limits for the different clouds). We also assume the same temperature for starless and protostellar cores, thereby, we do not consider the possibility of internal heating due to a central protostar. From Equations \ref{jeansEq} and \ref{massEq}, \MMJ\ decreases for higher temperatures. Thus, if protostellar cores are intrinsically warmer than starless cores, then we are overestimating our protostellar \MMJ\ ratios. 

We adopted temperatures from ammonia surveys published by \citet[for Ophiuchus]{Friesen09}, \citet[for Taurus]{Pratap97}, \citet[for Perseus]{Rosolowsky08}, and \citet[for Orion]{Wilson99}. 
For the cores in our sample with ammonia-derived temperatures, we compared the average temperature between the starless and protostellar populations. Generally, the temperatures of the two populations agreed within 1 K. In Perseus, for example, we found an average starless core temperature of $12.5 \pm 2.7$ K with 39 starless cores and an average protostellar core temperature of $12.2 \pm 1.6$ K with 35 protostellar cores. Unfortunately, the other clouds had far fewer cores with ammonia-derived temperatures. For example, we found only 9 starless cores and 2 protostellar cores listed in the survey by \citet{Friesen09} for Ophiuchus, 1 starless core and 1 protostellar core in \citet{Benson89}\footnote{ \citet{Pratap97} only published the average ammonia-derived temperature for their entire survey and not the temperatures from individual cores.} for Taurus, and 7 starless cores and 3 protostellar cores in \citet{Wilson99} for Orion. 

Since the two core populations gave similar temperatures, we adopted a single temperature for both starless and protostellar cores based on the average core temperature across the entire cloud (see Table 1). Nevertheless, the uncertainty in core temperature remains a moderate concern. Ideally, we would want accurate temperatures for each individual core, e.g., through detailed ammonia observations or with \emph{Herschel} observations.

\subsection{Re-examining the Super-Jeans Starless Cores}\label{reexamine}

There were 729 dense cores identified as starless or protosellar in \citetalias{Sadavoy10} for our clouds. The majority of these dense cores are sub-critical ($\MMJ < 1$). Of the 555 starless cores in \citetalias{Sadavoy10}, only 17 had \MMJ\ ratios greater than 2 (for Ophiuchus and Taurus) or 4.5 (for Perseus and Orion). We will not discuss cores with \MMJ\ values that fall below these super-critical ratios.

Although we find 17 candidate starless super-Jeans cores, it is possible these cores are misidentified protostellar cores. In \citetalias{Sadavoy10}, YSO candidates were selected based on specific colors and firm \emph{Spitzer} detections. 
To determine if our 17 candidate starless super-Jeans cores are really starless, we examined Spitzer images and the literature. The results are given in Table \ref{newClassTable} and discussed below.

\subsubsection{Starless Super-Jeans Cores}

Of our 17 super-Jeans starless candidates, we found that 3 appear to be genuinely starless in the Spitzer maps (see Figures \ref{SpitzerFig}a,b,c). These are Oph-2 ($\MMJ = 2.05$), Per-2 ($\MMJ = 4.79$), and Per-6 ($\MMJ = 4.94$). For these cores, there are no infrared sources towards them in any of the Spitzer bands. Therefore, these cores represent an unusual physical state and should be interesting targets for future study. 

In the literature, however, these three cores have been cited as protostellar. For example, \citet{Jorgensen08} considered Oph-2 to be protostellar using a measure of the core concentration ($C=0.74$) and \citet{Nutter06} considered Oph-2 to be protostellar due to the proximity of the infrared source IRAS 16293-2442E. Surprisingly, there is no \emph{Spitzer} emission corresponding to IRAS 16293-2442E. Similarly, Per-2 and Per-6 have been labeled as protostellar (see \citealt{Hatchell07} and \citealt{Enoch08}), though we do not observe any infrared sources towards these cores at any of the Spitzer wavelengths. Thus, these three cores appear to be starless cores with masses that well exceed their predicted Jeans mass by factors of 2-5.

\subsubsection{Misidentified Protostellar Cores}


Of our 17 candidate starless super-Jeans cores, only 3 (Tau-1, Tau-2, and Per-7) were re-classified as protostellar. For Taurus, Tau-1 and Tau-2 are located in L1551, which was not observed by the Taurus Spitzer Survey team (see Padgett et al. 2010, in prep). Infrared maps of L1551 by Fazio et al. (taken from the \emph{Spitzer} archive) revealed compact infrared emission across all wavelengths and asymmetric nebulosity towards both cores (see Figure \ref{SpitzerFig}d). Both cores are associated with X-ray emission suggesting accretion (\citealt{Gudel07}) as well as associated outflows and reflection nebulae (\citealt{HayashiPyo09}). Similarly, Per-7 was misidentified as starless due to insufficient infrared observations towards the core (Per-7 is the deeply embedded Class 0 source, IRAS 4A). Per-7 was misidentified as starless in \citetalias{Sadavoy10} due to poor IRAC detections (see lack of 8.0 \um\ emission in Figure \ref{SpitzerFig}c) towards the core. As a result, Per-7 failed the color criteria in \citetalias{Sadavoy10} and was classified as starless.  

\subsubsection{Undetermined Sources}

Most of our candidate starless super-Jeans cores are located in clustered environments (e.g., all five candidates in Orion are in OMC-1). The infrared emission towards cores in crowded environments can be obscured, such as from infrared emission of nearby, more evolved YSOs. As such, faint protostellar emission might be lost, contaminated or misinterpreted (i.e., emission from outflow shocks could be mistaken for protostellar signatures; see \citealt{Hatchell09}).

We identified 11 of our 17 candidate cores as having ``undetermined'' classifications, where the presence or absence of protostellar associations was ambiguous. These undetermined cores include one core in Ophiuchus and five each in Perseus and Orion. For Orion, there was significant large-scale nebulosity observed towards each of the five cores. Therefore, any faint infrared emission from a protostar could be easily lost by this large-scale nebulosity. For the undetermined sources in Ophiuchus and Perseus, there were nearby bright infrared sources. For example, Figure \ref{SpitzerFig}e shows the 8 \um\ map around Oph-1. There is no obvious compact infrared emission towards Oph-1, however, the core is located near a bright infrared object, possibly the T Tauri star BF-10 (\citealt{Imanishi03}). Only Per-8 appears to have a compact infrared source towards it (see Figure \ref{SpitzerFig}f), however, this infrared source has fairly blue colors and would fail the YSO-selecting color criteria from \citet{Evans09}, \citet{Megeath09}, and \citetalias{Sadavoy10}. Thus, the compact infrared source towards Per-8 could be a chance coincidence from an object external to the cloud.


\section{Discussion}\label{Discussion}

Table \ref{newClassTable} lists the new classifications for each of our candidate starless super-Jeans cores and Figure \ref{threeSymFigure} illustrates the \MMJ\ ratios using these final classifications. With the new classification system, most of the super-Jeans cores in each cloud are undetermined or protostellar. For example, all super-Jeans cores in Taurus are protostellar. In Ophiuchus and Perseus, there are a few starless super-Jeans cores near our imposed \MMJ\ limit, whereas the undetermined super-Jeans cores can have very large \MMJ\ ratios.

Super-Jeans cores can only be starless if they have non-thermal processes providing internal support. Therefore, these cores are excellent test cases for exploring these processes, such as rotation, turbulence, or magnetic fields. If non-thermal mechanisms play a significant role in the evolution of dense cores into stars, then it is important to understand when and how they influence their evolution. 

Alternatively, it is possible that these cores are unresolved and only appear massive because they are blends from several less massive cores (i.e., from line-of-sight coincidences or fragmentation). Since our effective SCUBA beam is $\sim 23\arcsec$, we cannot detect structure on scales smaller than 0.014 pc (for Ophiuchus) or 0.028 pc (for Perseus). In addition, many stars are multiples (\citealt{Duquennoy91}), so fragmentation in cores is expected, though when this process occurs is unknown. Dense cores, however, generally show narrow, single line emission compared to their parent cloud (e.g.,  \citealt{Tatematsu04}, \citealt{Kirk07}, \citealt{Friesen09}, \citealt{Pineda10}), thus it becomes unlikely that they are line-of-sight coincidences. As well, there are few observations of potential core fragmentation (e.g., \citealt{JKirk09}) and for the most part, dense cores appear to be smooth (see \citealt{Olmi05}, \citealt{Schnee10}). 

For the undetermined super-Jeans cores, most are found in highly crowded regions (i.e., NGC 1333, OMC-1), where crowding and highly energetic events (i.e., outflows) confuse protostellar emission and could result in undetermined classification. Determining whether these cores are starless or protostellar is still very important. If all the undetermined super-Jeans cores are starless, then massive starless cores may be linked to crowded environments. Such an observation would suggest that CMFs depend on environment. If all the undetermined super-Jeans cores are protostellar, then massive cores tend to be evolved (as proposed before, e.g., see \citealt{Enoch08}). In this case, CMFs would lack the massive core tail required to reproduce the IMF.

We caution that in these very crowded environments, the temperature may be higher than in more isolated regions. There is a large temperature dependence in the \MMJ\ ratio (see Equations \ref{jeansEq} and \ref{massEq}). If the undetermined super-Jeans cores have underestimated temperatures, then we are overestimating the \MMJ\ ratio. In recent study of Perseus by \citet{Schnee09}, however, the starless core temperatures were relatively constant (within a few Kelvin) over the cloud. Future study with \emph{Herschel} should reveal the temperature structure in these dense cores.

\section{Conclusions}\label{conc}

From a sample of 555 starless dense cores (see \citetalias{Sadavoy10}), we identified 17 cores with masses that exceeded their expected Jeans mass. We examined the classification of these 17 candidate starless super-Jeans cores more closely using Spitzer maps at 3.6-24 $\mu$m and only three (Oph-2, Per-2, and Per-6) maintained their classification as starless super-Jeans cores based on no apparent infrared emission towards the location of the corresponding cores. These 3 starless super-Jeans cores represent excellent observational targets for future studies. Starless super-Jeans cores should be thermally unstable, requiring additional non-thermal supports to prevent collapse. Therefore, these three cores might represent the cusp of the pre-stellar stage just prior to the onset of star formation.

In addition, we labeled 11 of our 17 candidate cores with an ``undetermined'' classification (see Table \ref{newClassTable}), due to ambiguous infrared emission towards the core. These undetermined super-Jeans cores are also very interesting targets for follow-up study since their classifications could point to a mass bias in the CMF, which in turn, could affect the observed relation between the CMF and the IMF. Finally, of these 17 candidates, we re-classified 3 objects as protostellar (these 3 cores were misidentified as starless in \citetalias{Sadavoy10} due to incomplete or undetected fluxes in the IRAC bands). 

Here, we have created a list of interesting targets for follow-up observations both photometrically (i.e., with interferometric continuum observations) or spectroscopically (i.e., with tracers for infall or rotation). For example, we can use high resolution interferometric observations to distinguish between an unstable, centrally peaked density structure and a smooth, stable density profile. Such observations could be very useful in constraining simulations of dense core collapse and formation. Our starless super-Jeans cores may represent the strongest cases of significant non-thermal supports and our undetermined super-Jeans cores could represent a mass bias in CMFs. Therefore, further study of these few starless and undetermined super-Jeans cores (out of hundreds of cores in these regions) is important to understand the processes involved in star formation.

\acknowledgments{The authors thank the anonymous referee for comments that improved this work. This work was possible with funding from a Natural Sciences and Engineering Research Council of Canada CGS award and a Discovery Grant. The James Clerk Maxwell Telescope is operated by The Joint Astronomy Centre on behalf of the Science and Technology Facilities Council of the United Kingdom, the Netherlands Organisation for Scientific Research, and the National Research Council of Canada. This work is based in part on observations made with the Spitzer Space Telescope, which is operated by the Jet Propulsion Laboratory, California Institute of Technology under a contract with NASA. 

\bibliographystyle{apj}
\bibliography{references}

\begin{thebibliography}{47}
\expandafter\ifx\csname natexlab\endcsname\relax\def\natexlab#1{#1}\fi

\bibitem[{{Bate}(2009)}]{Bate09}
{Bate}, M.~R. 2009, \mnras, 392, 1363

\bibitem[{{Benson} \& {Myers}(1989)}]{Benson89}
{Benson}, P.~J., \& {Myers}, P.~C. 1989, \apjs, 71, 89

\bibitem[{{Bonnor}(1956)}]{Bonnor56}
{Bonnor}, W.~B. 1956, \mnras, 116, 351

\bibitem[{{Clark} {et~al.}(2008){Clark}, {Bonnell}, \& {Klessen}}]{Clark08}
{Clark}, P.~C., {Bonnell}, I.~A., \& {Klessen}, R.~S. 2008, \mnras, 386, 3

\bibitem[{{Di Francesco} {et~al.}(2007){Di Francesco}, {Evans}, {Caselli},
  {Myers}, {Shirley}, {Aikawa}, \& {Tafalla}}]{difran07}
{Di Francesco}, J., {Evans}, II, N.~J., {Caselli}, P., {Myers}, P.~C.,
  {Shirley}, Y., {Aikawa}, Y., \& {Tafalla}, M. 2007, in Protostars and Planets
  V, ed. B.~{Reipurth}, D.~{Jewitt}, \& K.~{Keil}, 17--32

\bibitem[{{Di Francesco} {et~al.}(2008){Di Francesco}, {Johnstone}, {Kirk},
  {MacKenzie}, \& {Ledwosinska}}]{difran08}
{Di Francesco}, J., {Johnstone}, D., {Kirk}, H., {MacKenzie}, T., \&
  {Ledwosinska}, E. 2008, \apjs, 175, 277

\bibitem[{{Duquennoy} \& {Mayor}(1991)}]{Duquennoy91}
{Duquennoy}, A., \& {Mayor}, M. 1991, \aap, 248, 485

\bibitem[{{Ebert}(1955)}]{Ebert55}
{Ebert}, R. 1955, Zeitschrift fur Astrophysik, 37, 217

\bibitem[{{Enoch} {et~al.}(2009){Enoch}, {Evans}, {Sargent}, \&
  {Glenn}}]{Enoch09}
{Enoch}, M.~L., {Evans}, II, N.~J., {Sargent}, A.~I., \& {Glenn}, J. 2009,
  \apj, 692, 973

\bibitem[{{Enoch} {et~al.}(2008){Enoch}, {Evans}, {Sargent}, {Glenn},
  {Rosolowsky}, \& {Myers}}]{Enoch08}
{Enoch}, M.~L., {Evans}, II, N.~J., {Sargent}, A.~I., {Glenn}, J.,
  {Rosolowsky}, E., \& {Myers}, P. 2008, \apj, 684, 1240

\bibitem[{{Evans} {et~al.}(2009){Evans}, {Dunham}, {J{\o}rgensen}, {Enoch},
  {Mer{\'{\i}}n}, {van Dishoeck}, {Alcal{\'a}}, {Myers}, {Stapelfeldt},
  {Huard}, {Allen}, {Harvey}, {van Kempen}, {Blake}, {Koerner}, {Mundy},
  {Padgett}, \& {Sargent}}]{Evans09}
{Evans}, II, N.~J., {Dunham}, M.~M., {J{\o}rgensen}, J.~K., {Enoch}, M.~L.,
  {Mer{\'{\i}}n}, B., {van Dishoeck}, E.~F., {Alcal{\'a}}, J.~M., {Myers},
  P.~C., {Stapelfeldt}, K.~R., {Huard}, T.~L., {Allen}, L.~E., {Harvey}, P.~M.,
  {van Kempen}, T., {Blake}, G.~A., {Koerner}, D.~W., {Mundy}, L.~G.,
  {Padgett}, D.~L., \& {Sargent}, A.~I. 2009, \apjs, 181, 321

\bibitem[{{Friesen} {et~al.}(2009){Friesen}, {Di Francesco}, {Shirley}, \&
  {Myers}}]{Friesen09}
{Friesen}, R.~K., {Di Francesco}, J., {Shirley}, Y.~L., \& {Myers}, P.~C. 2009,
  \apj, 697, 1457

\bibitem[{{Goldsmith} {et~al.}(2008){Goldsmith}, {Heyer}, {Narayanan}, {Snell},
  {Li}, \& {Brunt}}]{Goldsmith08}
{Goldsmith}, P.~F., {Heyer}, M., {Narayanan}, G., {Snell}, R., {Li}, D., \&
  {Brunt}, C. 2008, \apj, 680, 428

\bibitem[{{G{\"u}del} {et~al.}(2007){G{\"u}del}, {Briggs}, {Arzner}, {Audard},
  {Bouvier}, {Feigelson}, {Franciosini}, {Glauser}, {Grosso}, {Micela},
  {Monin}, {Montmerle}, {Padgett}, {Palla}, {Pillitteri}, {Rebull}, {Scelsi},
  {Silva}, {Skinner}, {Stelzer}, \& {Telleschi}}]{Gudel07}
{G{\"u}del}, M., {Briggs}, K.~R., {Arzner}, K., {Audard}, M., {Bouvier}, J.,
  {Feigelson}, E.~D., {Franciosini}, E., {Glauser}, A., {Grosso}, N., {Micela},
  G., {Monin}, J., {Montmerle}, T., {Padgett}, D.~L., {Palla}, F.,
  {Pillitteri}, I., {Rebull}, L., {Scelsi}, L., {Silva}, B., {Skinner}, S.~L.,
  {Stelzer}, B., \& {Telleschi}, A. 2007, \aap, 468, 353

\bibitem[{{Hatchell} \& {Dunham}(2009)}]{Hatchell09}
{Hatchell}, J., \& {Dunham}, M.~M. 2009, \aap, 502, 139

\bibitem[{{Hatchell} {et~al.}(2007){Hatchell}, {Fuller}, {Richer}, {Harries},
  \& {Ladd}}]{Hatchell07}
{Hatchell}, J., {Fuller}, G.~A., {Richer}, J.~S., {Harries}, T.~J., \& {Ladd},
  E.~F. 2007, \aap, 468, 1009

\bibitem[{{Hayashi} \& {Pyo}(2009)}]{HayashiPyo09}
{Hayashi}, M., \& {Pyo}, T. 2009, \apj, 694, 582

\bibitem[{{Henning} {et~al.}(1995){Henning}, {Michel}, \&
  {Stognienko}}]{Henning95}
{Henning}, T., {Michel}, B., \& {Stognienko}, R. 1995, \planss, 43, 1333

\bibitem[{{Imanishi} {et~al.}(2003){Imanishi}, {Nakajima}, {Tsujimoto},
  {Koyama}, \& {Tsuboi}}]{Imanishi03}
{Imanishi}, K., {Nakajima}, H., {Tsujimoto}, M., {Koyama}, K., \& {Tsuboi}, Y.
  2003, \pasj, 55, 653

\bibitem[{{Johnstone} {et~al.}(2001){Johnstone}, {Fich}, {Mitchell}, \&
  {Moriarty-Schieven}}]{Johnstone01}
{Johnstone}, D., {Fich}, M., {Mitchell}, G.~F., \& {Moriarty-Schieven}, G.
  2001, \apj, 559, 307

\bibitem[{{Johnstone} {et~al.}(2000){Johnstone}, {Wilson}, {Moriarty-Schieven},
  {Joncas}, {Smith}, {Gregersen}, \& {Fich}}]{Johnstone00}
{Johnstone}, D., {Wilson}, C.~D., {Moriarty-Schieven}, G., {Joncas}, G.,
  {Smith}, G., {Gregersen}, E., \& {Fich}, M. 2000, \apj, 545, 327

\bibitem[{{J{\o}rgensen} {et~al.}(2007){J{\o}rgensen}, {Johnstone}, {Kirk}, \&
  {Myers}}]{Jorgensen07}
{J{\o}rgensen}, J.~K., {Johnstone}, D., {Kirk}, H., \& {Myers}, P.~C. 2007,
  \apj, 656, 293

\bibitem[{{J{\o}rgensen} {et~al.}(2008){J{\o}rgensen}, {Johnstone}, {Kirk},
  {Myers}, {Allen}, \& {Shirley}}]{Jorgensen08}
{J{\o}rgensen}, J.~K., {Johnstone}, D., {Kirk}, H., {Myers}, P.~C., {Allen},
  L.~E., \& {Shirley}, Y.~L. 2008, \apj, 683, 822

\bibitem[{{Kirk} {et~al.}(2007){Kirk}, {Johnstone}, \& {Tafalla}}]{Kirk07}
{Kirk}, H., {Johnstone}, D., \& {Tafalla}, M. 2007, \apj, 668, 1042

\bibitem[{{Kirk} {et~al.}(2009){Kirk}, {Ward-Thompson}, {Di Francesco},
  {Bourke}, {Evans}, {Mer{\'{\i}}n}, {Allen}, {Cieza}, {Dunham}, {Harvey},
  {Huard}, {J{\o}rgensen}, {Miller}, {Noriega-Crespo}, {Peterson}, {Ray}, \&
  {Rebull}}]{JKirk09}
{Kirk}, J.~M., {Ward-Thompson}, D., {Di Francesco}, J., {Bourke}, T.~L.,
  {Evans}, II, N.~J., {Mer{\'{\i}}n}, B., {Allen}, L.~E., {Cieza}, L.~A.,
  {Dunham}, M.~H., {Harvey}, P., {Huard}, T., {J{\o}rgensen}, J.~K., {Miller},
  J.~F., {Noriega-Crespo}, A., {Peterson}, D., {Ray}, T.~P., \& {Rebull}, L.~M.
  2009, ArXiv astro-ph/0903.4063

\bibitem[{{Kunz} \& {Mouschovias}(2009)}]{Kunz09}
{Kunz}, M.~W., \& {Mouschovias}, T.~C. 2009, \mnras, 399, L94

\bibitem[{{Lada} {et~al.}(2008){Lada}, {Muench}, {Rathborne}, {Alves}, \&
  {Lombardi}}]{Lada08}
{Lada}, C.~J., {Muench}, A.~A., {Rathborne}, J., {Alves}, J.~F., \& {Lombardi},
  M. 2008, \apj, 672, 410

\bibitem[{{Larson}(2003)}]{Larson03}
{Larson}, R.~B. 2003, Reports on Progress in Physics, 66, 1651

\bibitem[{{McKee} \& {Ostriker}(2007)}]{McKee07}
{McKee}, C.~F., \& {Ostriker}, E.~C. 2007, \araa, 45, 565

\bibitem[{{Megeath} {et~al.}(2009){Megeath}, {Allgaier}, {Young}, {Allen},
  {Pipher}, \& {Wilson}}]{Megeath09}
{Megeath}, S.~T., {Allgaier}, E., {Young}, E., {Allen}, T., {Pipher}, J.~L., \&
  {Wilson}, T.~L. 2009, \aj, 137, 4072

\bibitem[{{Motte} {et~al.}(1998){Motte}, {Andre}, \& {Neri}}]{Motte98}
{Motte}, F., {Andre}, P., \& {Neri}, R. 1998, \aap, 336, 150

\bibitem[{{Myers}(1999)}]{Myers99}
{Myers}, P.~C. 1999, in The Physics and Chemistry of the Interstellar Medium,
  ed. {V.~Ossenkopf, J.~Stutzki, \& G.~Winnewisser}, 227

\bibitem[{{Nutter} {et~al.}(2008){Nutter}, {Kirk}, {Stamatellos}, \&
  {Ward-Thompson}}]{Nutter08}
{Nutter}, D., {Kirk}, J.~M., {Stamatellos}, D., \& {Ward-Thompson}, D. 2008,
  \mnras, 384, 755

\bibitem[{{Nutter} {et~al.}(2006){Nutter}, {Ward-Thompson}, \&
  {Andr{\'e}}}]{Nutter06}
{Nutter}, D., {Ward-Thompson}, D., \& {Andr{\'e}}, P. 2006, \mnras, 368, 1833

\bibitem[{{Olmi} {et~al.}(2005){Olmi}, {Testi}, \& {Sargent}}]{Olmi05}
{Olmi}, L., {Testi}, L., \& {Sargent}, A.~I. 2005, \aap, 431, 253

\bibitem[{{Peterson} \& {Megeath}(2008)}]{PetersonMegeath08}
{Peterson}, D.~E., \& {Megeath}, S.~T. 2008, {The Orion Molecular Cloud 2/3 and
  NGC 1977 Regions} (Handbook of Star Forming Regions, Volume I: The Northern
  Sky ASP Monograph Publications, Vol.~4.~Edited by Bo Reipurth, p.590-620)

\bibitem[{{Pineda} {et~al.}(2010){Pineda}, {Goodman}, {Arce}, {Caselli},
  {Foster}, {Myers}, \& {Rosolowsky}}]{Pineda10}
{Pineda}, J.~E., {Goodman}, A.~A., {Arce}, H.~G., {Caselli}, P., {Foster},
  J.~B., {Myers}, P.~C., \& {Rosolowsky}, E.~W. 2010, \apjl, 712, L116

\bibitem[{{Pratap} {et~al.}(1997){Pratap}, {Dickens}, {Snell}, {Miralles},
  {Bergin}, {Irvine}, \& {Schloerb}}]{Pratap97}
{Pratap}, P., {Dickens}, J.~E., {Snell}, R.~L., {Miralles}, M.~P., {Bergin},
  E.~A., {Irvine}, W.~M., \& {Schloerb}, F.~P. 1997, \apj, 486, 862

\bibitem[{{Rosolowsky} {et~al.}(2008){Rosolowsky}, {Pineda}, {Foster},
  {Borkin}, {Kauffmann}, {Caselli}, {Myers}, \& {Goodman}}]{Rosolowsky08}
{Rosolowsky}, E.~W., {Pineda}, J.~E., {Foster}, J.~B., {Borkin}, M.~A.,
  {Kauffmann}, J., {Caselli}, P., {Myers}, P.~C., \& {Goodman}, A.~A. 2008,
  \apjs, 175, 509

\bibitem[{{Sadavoy} {et~al.}(2010){Sadavoy}, {Di Francesco}, {Bontemps},
  {Megeath}, {Rebull}, {Allgaier}, {Carey}, {Gutermuth}, {Hora}, {Huard},
  {McCabe}, {Muzerolle}, {Noriega-Crespo}, {Padgett}, \& {Terebey}}]{Sadavoy10}
{Sadavoy}, S.~I., {Di Francesco}, J., {Bontemps}, S., {Megeath}, S.~T.,
  {Rebull}, L.~M., {Allgaier}, E., {Carey}, S., {Gutermuth}, R., {Hora}, J.,
  {Huard}, T., {McCabe}, C., {Muzerolle}, J., {Noriega-Crespo}, A., {Padgett},
  D., \& {Terebey}, S. 2010, \apj, 710, 1247

\bibitem[{{Schnee} {et~al.}(2010){Schnee}, {Enoch}, {Johnstone}, {Culverhouse},
  {Leitch}, {Marrone}, \& {Sargent}}]{Schnee10}
{Schnee}, S., {Enoch}, M., {Johnstone}, D., {Culverhouse}, T., {Leitch}, E.,
  {Marrone}, D.~P., \& {Sargent}, A. 2010, ArXiv e-prints astro-ph 1005.5169

\bibitem[{{Schnee} {et~al.}(2009){Schnee}, {Rosolowsky}, {Foster}, {Enoch}, \&
  {Sargent}}]{Schnee09}
{Schnee}, S., {Rosolowsky}, E., {Foster}, J., {Enoch}, M., \& {Sargent}, A.
  2009, \apj, 691, 1754

\bibitem[{{Shu} {et~al.}(1987){Shu}, {Adams}, \& {Lizano}}]{Shu87}
{Shu}, F.~H., {Adams}, F.~C., \& {Lizano}, S. 1987, \araa, 25, 23

\bibitem[{{Tatematsu} {et~al.}(2004){Tatematsu}, {Umemoto}, {Kandori}, \&
  {Sekimoto}}]{Tatematsu04}
{Tatematsu}, K., {Umemoto}, T., {Kandori}, R., \& {Sekimoto}, Y. 2004, \apj,
  606, 333

\bibitem[{{Ward-Thompson} {et~al.}(2007){Ward-Thompson}, {Andr{\'e}},
  {Crutcher}, {Johnstone}, {Onishi}, \& {Wilson}}]{Ward-T07}
{Ward-Thompson}, D., {Andr{\'e}}, P., {Crutcher}, R., {Johnstone}, D.,
  {Onishi}, T., \& {Wilson}, C. 2007, in Protostars and Planets V, ed.
  B.~{Reipurth}, D.~{Jewitt}, \& K.~{Keil}, 33--46

\bibitem[{{Williams} {et~al.}(1994){Williams}, {de Geus}, \&
  {Blitz}}]{Clumpfind}
{Williams}, J.~P., {de Geus}, E.~J., \& {Blitz}, L. 1994, \apj, 428, 693

\bibitem[{{Wilson} {et~al.}(1999){Wilson}, {Mauersberger}, {Gensheimer},
  {Muders}, \& {Bieging}}]{Wilson99}
{Wilson}, T.~L., {Mauersberger}, R., {Gensheimer}, P.~D., {Muders}, D., \&
  {Bieging}, J.~H. 1999, \apj, 525, 343

\end{thebibliography}


\pagebreak

\begin{figure}[h!]
\includegraphics[scale=0.779]{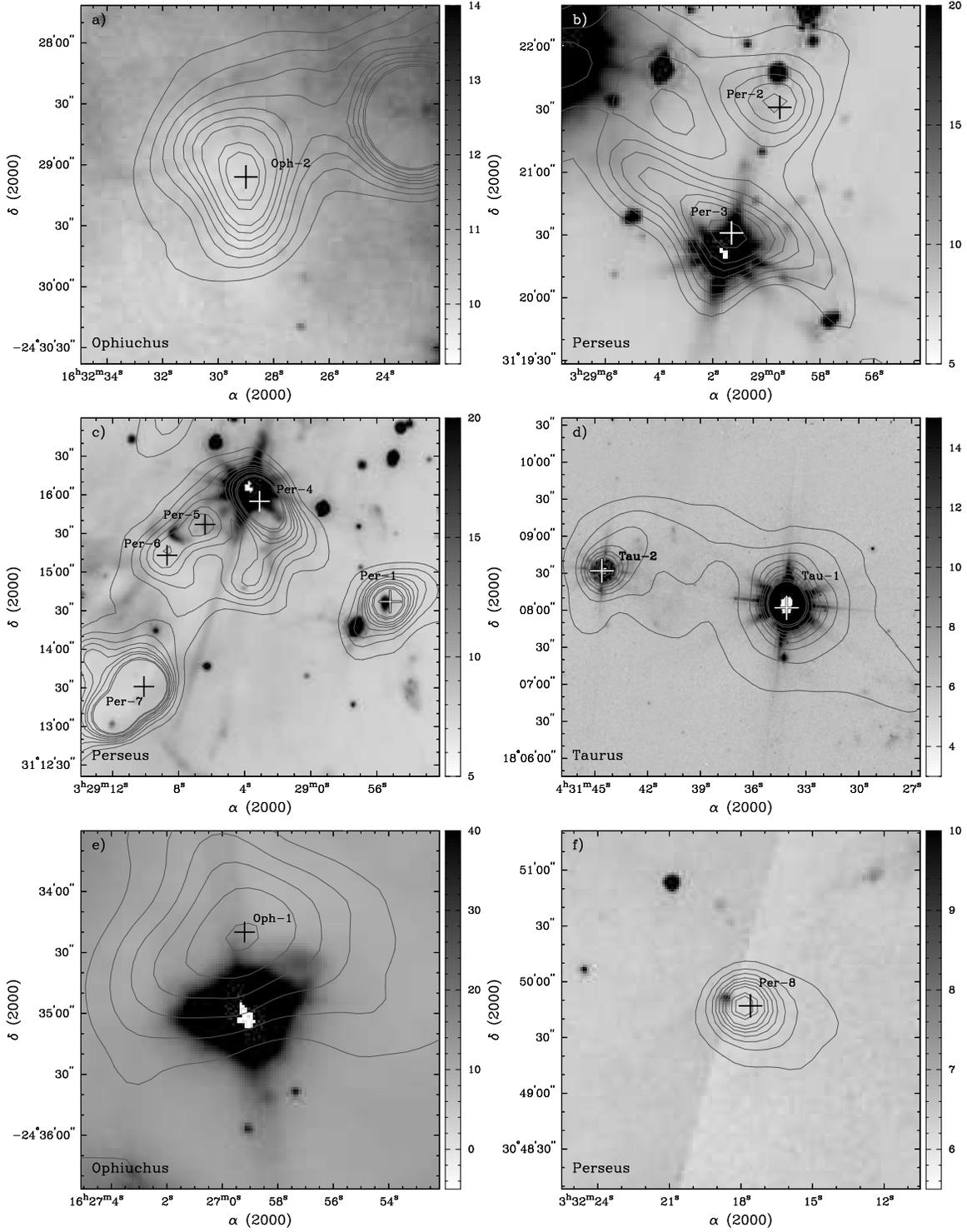}\\[-8mm]
\caption{Spitzer 8 \um\ images of super-Jeans cores in Ophiuchus, Taurus, and Perseus. Crosses denote the locations of peak 850 \um\ intensity for each respective core and the contours denote 850 \um\ fluxes from 0.2 - 2.0 \Jybeam\ in steps of 0.2 \Jybeam.\label{SpitzerFig}}
\end{figure}

\pagebreak

\begin{figure}[h!]
\includegraphics[scale=0.9]{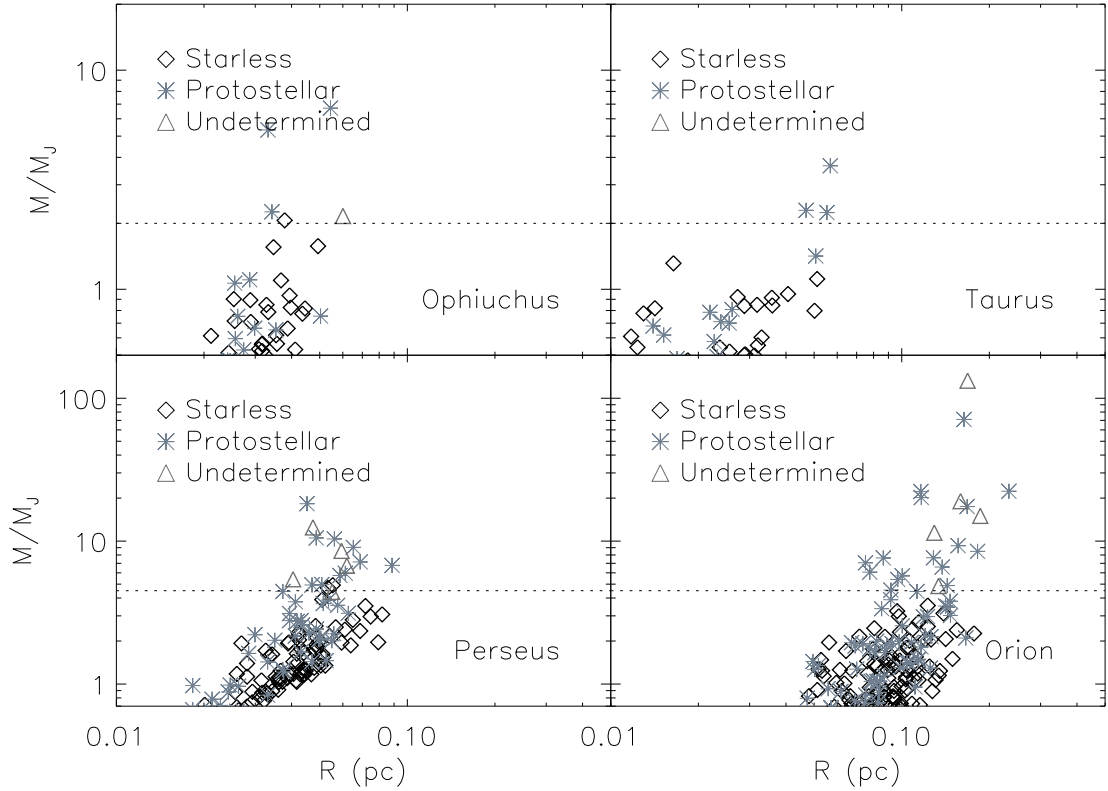}
\caption{Comparison of core mass, scaled by the Jeans mass, with observed core size for our starless, undetermined, and protostellar cores (as classified here). Masses were calculated using the temperatures found in Table \ref{coreTable}. The dashed lines indicate \MMJ\ $= 2$ for Ophiuchus and Taurus and \MMJ\ $= 4.5$ for Perseus and Orion, respectively.\label{threeSymFigure}}
\end{figure}

\pagebreak


\begin{table}[h!]
\caption{Starless and Protostellar Cores in Gould Belt Clouds\label{coreTable}}
\begin{tabular}{lcccc}
\hline
\multirow{2}{*}{Cloud} & Temperature\tablenotemark{a} & Distance\tablenotemark{b} & Starless & Protostellar\\
 & (K) & (pc) & Cores & Cores\\
\hline\hline
Ophiuchus & 15 & 125 & 97 & 27\\
Taurus & 13 & 140 & 69 & 18\\
Perseus & 11 & 250 & 97 & 46\\
Orion & 20 & 450 & 292 & 83\\
\hline
\end{tabular}
\tablenotetext{a}{References for temperature are \citet{Friesen09}, \citet{Pratap97}, \citet{Rosolowsky08}, and \citet{Wilson99}, respectively.}
\tablenotetext{b}{References for distance are \citet{Enoch09}, \citet{Goldsmith08}, \citet{Enoch09}, and \citet{PetersonMegeath08}, respectively.}
\end{table}

\begin{table}[h!]
\caption{Candidate Starless Super-Jeans Cores\tablenotemark{a}\label{newClassTable}}
\begin{tabular}{lllcccl}
\hline
No. & SCUBA Core & Nearby Sources\tablenotemark{b}& Mass & $R_{eff}$ & \MMJ\ & New Classification\tablenotemark{c}\\
& (J2000.0) & & ($M_{\odot}$) & (pc) & & \\
\hline\hline
Oph-1 & J162659.2$-$243420 & Oph C-MM3 & 5.3 & 0.060 & 2.2 & undetermined\\
Oph-2 & J163229.0$-$242906 & IRAS 16293-2422E & 3.2 & 0.038 & 2.1 & starless\\
\hline
Tau-1 & J043134.1+180802 & LDN 1551 IRS 5 & 7.3 & 0.057 & 3.6 & protostellar\\
Tau-2 & J043144.6+180832 & LDN 1551NE & 3.8 & 0.047 & 2.3 & protostellar\\
\hline
Per-1 & J032855.2+311437 & IRAS 2A & 12 & 0.062 & 6.7 & undetermined\\
Per-2 & J032859.5+312131 & $\cdots$ & 7.6 & 0.053 & 4.8 & starless\\
Per-3 & J032901.3+312031 & IRAS 6 & 15 & 0.059 & 8.6 & undetermined\\
Per-4 & J032903.1+311555 & NGC 1333 13A & 18 & 0.047 & 13 & undetermined\\
Per-5 & J032906.4+311537 & HH 8 & 6.5 & 0.040 & 5.5 & undetermined\\
Per-6 & J032908.7+311513 & $\cdots$& 8.1 & 0.055 & 4.9 & starless\\
Per-7 & J032910.1+311331 & IRAS 4A & 25 & 0.045 & 18 & protostellar\\
Per-8 & J033217.6+304947 & IRAS 03292+3039 & 7.2 & 0.055 & 4.4 & undetermined\\
\hline
Ori-1 & J053514.4$-$052232 & SMA 1 & 1200 & 0.17 & 130 & undetermined\\
Ori-2 & J053514.4$-$052608 & COUP 617  & 150 & 0.19 & 15 & undetermined\\
Ori-3 & J053516.8$-$051926 & OMC-N4 & 160 & 0.16 & 19 & undetermined\\
Ori-4 & J053522.0$-$052508 & COUP 1194  & 80 & 0.13 & 11 & undetermined\\
Ori-5 & J053524.8$-$052220 & COUP 1314 & 35 & 0.13 & 5.0 & undetermined\\
\hline
\end{tabular}
\tablenotetext{a}{Super-Jeans cores were defined based on $\MMJ > 2$ for Ophiuchus and Taurus and $\MMJ > 4.5$ for Perseus and Orion.}
\tablenotetext{b}{Nearby sources were obtained using SIMBAD.}
\tablenotetext{c}{See \S \ref{reexamine} for explanations.}
\end{table}

\end{document}